\documentstyle[12pt]{article}
\renewcommand
\baselinestretch{1.3}

\begin{document}

\def\beq{\begin{equation}}
\def\eeq{\end{equation}}
\def\bce{\begin{center}}
\def\ece{\end{center}}
\def\bea{\begin{eqnarray}}
\def\eea{\end{eqnarray}}
\def\ben{\begin{enumerate}}
\def\een{\end{enumerate}}
\def\ul{\underline}
\def\ni{\noindent}
\def\nn{\nonumber}
\def\bs{\bigskip}
\def\ms{\medskip}
\def\tr{\mbox{Tr}\ }
\def\wt{\widetilde}
\def\rl{$\Real$}
\def\tu{\bigtriangleup}
\def\td{\bigtriangledown}
\def\brr{\begin{array}}
\def\err{\end{array}}
\def\ri{{\rm i}}
\def\nabar{\not\!\nabla}
\def\nabar2{{\not\!\nabla}^2}
\def\nablap{\frac{\not\nabla ^2}{\Lambda ^2}}
\def\nablam{\frac{\not\!\nabla ^2}{\Lambda ^2}}
\def\R{{\rm R}}

\def\ds{\displaystyle}
\def\Sp{{\rm Sp}\ }

\hfill UB-ECM-PF 94/33

\hfill hep-th/9411216

\hfill November 1994

\vspace*{3mm}

\begin{center}

{\LARGE \bf Higher derivative four-fermion model in curved
spacetime}

\vspace{4mm}

\renewcommand
\baselinestretch{0.8}
\medskip

\renewcommand
\baselinestretch{1.4}
{\sc E. Elizalde$^{a,b,}$}\footnote{E-mail: eli@zeta.ecm.ub.es},
{\sc S. Leseduarte$^{b,}$}\footnote{E-mail: lese@ecm.ub.es}, and
{\sc S.D. Odintsov$^{b,}$}\footnote{E-mail: odintsov@ecm.ub.es.
On leave from: Tomsk Pedagogical Institute, 634041 Tomsk, Russia}
\\
$^a$Center for Advanced Study CEAB, CSIC,
Cam\'{\i} de Sta. B\`arbara, 17300 Blanes\\
$^b$Department ECM and IFAE,
Faculty of Physics, University of  Barcelona, \\
Diagonal 647, 08028 Barcelona,
Catalonia, Spain \\

\vspace{5mm}

{\bf Abstract}

\end{center}

 We discuss the phase structure of a higher derivative four-fermion
model in four dimensions
in curved spacetime in frames of the $\frac{1}{N_c}$-expansion. First,
we
evaluate in our model the effective potential of two composite scalars
in the linear curvature approximation using a local momentum
representation in curved spacetime
for the higher-derivative propagator which naturally appears. The
symmetry breaking phenomenon and phase transition induced by curvature
are numerically investigated. A numerical study of the dynamically
generated fermionic mass, which depends on the coupling constants and on
the cuvature, is also presented.

\vspace{4mm}

\newpage

\ni {\bf 1. Introduction}.
The celebrated idea by Nambu \cite{1} to replace the fundamental
Higgs boson in the standard model (SM) by a top quark condensate, which
is responsible for dynamical symmetry breaking, has been investigated in
some detail in refs. \cite{2}. In frames of the Nambu-Jona-Lasinio
mechanism \cite{3}, and under some conditions, it was shown that in
the large-$N$ limit the physics of such Nambu-Jona-Lasinio (NJL) model
is equivalent to the physics of the SM with elementary scalar
fields.

 However, the original NJL model \cite{3} with gauge fields and
some of its modifications are considered now as non-renormalizable
effective theories where the presence of an ultraviolet cut-off
$\Lambda$ at the loop diagrams is a necessary condition. It is clear
that there are different possibilities to extend
the NJL model. Among such possibilities, a quite interesting one is the
introduction in the original Lagrangian of higher-derivative  terms in
the four fermion interaction \cite{4,5} or in the kinetic term
\cite{6}. It may be shown \cite{4} that the physics of such
generalized NJL models is still equivalent to the physics of the
SM. Moreover, the additional advantage appears, that the ultraviolet
cut-off $\Lambda $ shows up at the Lagrangian level.
That may help discuss the ambiguities which
turn out in the choice of a cut-off \cite{4,5,6} and violation
of background gauge invariance. From another side, the inclusion of
higher derivatives in effective theories gives the possibility to take
into account the structural effects of the medium and external fields,
like external gauge and gravitational fields or condensates.

 It is the purpose of this work to consider a higher derivative
four-fermion
model in curved spacetime and to study the influence of an external
gravitational field to the effective potential of composite scalars in
the large-$\frac{1}{N_c}$ expansion. In particular, we discuss the phase
structure and the dynamically generated fermion mass at non-zero
curvature.
This can be important in early Universe considerations. Notice that
similar questions for the standard NJL model in curved spacetime have
been already investigated in  refs. \cite{7,8,9} (for
a study of $\lambda \phi ^4 $ theory 
see \cite{11}). In particular, it was shown that
curvature may induce first order phase transitions \cite{8,9}. We
will here discuss  whether these considerations go over to the higher
derivative four fermion model at issue.
\ms

\ni {\bf 2. The model and its effective potential}.
We start by presenting the model which we set out to study
(it was introduced in \cite{5}).
Its Lagrangian density in curved spacetime (for an introduction to
quantum field theory in curved spacetime, see \cite{12} is given
by
\beq
{\cal L} = \bar{\psi} i \gamma ^\mu (x) \nabla _\mu \psi + \frac{1}{4
N_{c} \Lambda ^2} \left\{ \lambda _1 \left( \bar{\psi} \psi \right) ^2 +
3 \lambda _2 \left[ \bar{\psi} \left(1-2 \frac{\not\!\nabla ^2}{\Lambda
^2} \right) \psi \right]^2 \right\} \label{1} \ ,
\eeq
where $N_c$ is the number of fermionic species, $\gamma^\mu (x) $
the Dirac matrices in curved space, $\Lambda $ a cut-off parameter, and
$\lambda_1$ and $\lambda_2$ are coupling constants. We work in the
$\frac{1}{N_c}$-expansion scheme.

 By introducing some auxiliary fields $\chi _1$ and $\chi  _2$ we can
give a description of this nonrenormalizable theory by means of the
action
\beq
S = \int d^4 x \sqrt{g} \left\{ \bar{\psi} i \gamma ^\mu (x)
\nabla _\mu \psi - N_c \Lambda ^2 \left( \frac{\chi_1 ^2}{\lambda_1} +
\frac{\chi_2 ^2}{\lambda_2} \right) - \left[ \chi_1 \bar{\psi} \psi +
\sqrt{3} \chi_2 \bar{\psi} \left( 1- 2 \frac{\not\!\nabla ^2}{\Lambda
^2} \right) \psi \right]
\right\} \ .           \label{2}
\eeq

Our purpose is to study the influence of
external gravity on the dynamical
breaking and restoration patterns of the symmetry possessed by the
Lagrangian (\ref{1}), which is given by the transformations
$\psi \rightarrow \gamma _5 \psi  $ and $  \bar{\psi} \rightarrow
\bar{\psi} \gamma _5  \ .$
If we refer to the action (\ref{2}), the symmetry is realized by adding
the following transformations for the auxiliary fields:
$\chi_1 \rightarrow - \chi_1  $ and $  \chi_2 \rightarrow
-\chi_2$.

 The effective action of this model in the $N_c \rightarrow \infty$
limit is given by
\beq
V_{eff} = \Lambda ^2 \left( \frac{\chi _2 ^2}{\lambda _1} +
\frac{\chi_2 ^2}{\lambda _2} \right) + V_{1~eff} \ ,
\label{veff}
\eeq
with \[ V_{1~eff}=i \left( {\cal V}ol \right) ^{-1} \tr \ln
\left( i \not\!\nabla -m + 2 \sqrt{3} \chi_2 \frac{\not\!\nabla
^2}{\Lambda ^2} \right) \ ,\]
where $m=\chi_1 + \sqrt{3} \chi _2 \ .$
 Thus, to leading order in the $\frac{1}{N_c}$-expansion we have the
gap equations, as follows
\bea
\left. \frac{\partial V_{1 \ eff}}{\partial m}  \right| _{\chi _2} =
-i \left( {\cal V}ol \right) ^{-1} \tr  \frac{1}{i \not\!\nabla -m
+ 2 \sqrt{3} \chi _2 \frac{\not\nabla ^2}{\Lambda ^2}} \ ,
\label{deriv1}
\\ \left. \frac{\partial V_{1 \ eff}}{\partial {\chi _2}}  \right| _m =
i \left( {\cal V}ol \right) ^{-1} \tr  \frac{2 \sqrt{3}
\frac{\not\nabla ^2}{\Lambda ^2}}{i
\not\!\nabla
-m + 2 \sqrt{3} \chi _2 \frac{\not\nabla ^2}{\Lambda ^2}} \ .
\label{deriv2}
\eea
First of all, we have to calculate the fermionic propagator, which
satisfies \[ \left( i \not\!\nabla -m + 2 \sqrt{3} \chi _2
\frac{\not\!\nabla ^2}{\Lambda ^2} \right) {\cal G}(x,\,x')=\delta
(x,\,x') \ .\]
Once this propagator is obtained (apart from terms which disappear
under the $\tr $ operation) it will be immediate to produce
equations (\ref{deriv1}) and (\ref{deriv2}) in explicit form and also
the dependence of
the effective potential itself on the fields $\chi _1  $ and $  \chi
_2$. The details of this rather lengthy derivation are written in the
Appendix, where we give details about the calculation of the effective
potential up to terms linear in the curvature.

 The final outcome is shown below in natural variables, which are
obtained by using the following definitions
\beq
r=\frac{{\rm R}}{\Lambda ^2}\,, ~~~x_1=\frac{\chi _1}{\Lambda}\,
,~~~x_2= \frac{\chi_2}{\Lambda}\,,~~~a=x_1 + \sqrt{3} x_2\,
,~~~v=\frac{V_{eff}}{\Lambda ^4} \ .  \label{varnat}
\eeq
The `dimensionless' effective potential $v$ is given by
\bea
v & = & \frac{a^2}{\l_{1}}-\frac{a^2}{8 \pi ^2}+\frac{a^4}{32
\pi^2}-\frac{a^2 r}{96 \pi^2}-2 \sqrt{3} \frac{a x_{2}}{l_{1}}  \\ \nn
& & +\sqrt{3}\frac{a x_{2}}{4 \pi ^2} -\sqrt{3} \frac{a ^3 x_{2}}{2
\pi^2}+3
\frac{x_{2}^2}{l_{1}} + \frac{x_{2}^2}{l_{2}} -\frac{x_{2}^2}{2 \pi ^2} + 9
\frac{a ^2
x_{2} ^2}{4 \pi^2}+  \\  \nn
& & +\frac{r x_{2}^2}{16 \pi ^2} - 2 \sqrt{3} \frac{a x_{2} ^3}{\pi
^2}+9 \frac{x_{2} ^4}{4 \pi ^2}
 -\frac{a ^4 \ln  a^2  }{16\pi ^2}- \frac{a^2 r \ln
a^2  }{ 96 \pi^2}  \ . \label{potnat}
\eea
Note that in deriving  this expression we have taken into account only
terms up to quartic order on the fields $a$ and $x_2$.
The gap equations ---which one obtains by differentiating $v$ with
respect
to $x_1$ and $x_2$ and equating the results to zero--- are, respectively
\bea
x_{1} \left( 1 - {{8\,{{\pi }^2}}\over {l_{1}}} \right) & = &
  -2\,{3^{{3\over 2}}}\,{{x_{1}}^2}\,x_{2} -
  18\,x_{1}\,{{x_{2}}^2} - 8\,{\sqrt{3}}\,{{x_{2}}^3} \\ \nn
& & -{a^3}\,\ln a^2 - r\,\left( {{a}\over 6} +
     {{a\,\ln a^2}\over {12}} \right)  \ ,   \\
 x_2 \left( 1 - {{8\,\,{{\pi }^2}}\over {l_{2}}} \right) & = &
-2 \,{\sqrt{3}}\,{x_{1}}^3 -18\,{{x_{1}}^2} \,x_{2} -
  8\,{3^{{3\over 2}}}\,x_{1}\,{{x_{2}}^2} - 24\,{{x_{2}}^3}  \\ \nn
& & - {\sqrt{3}}\,{a^3}\,\ln  a^2
  -r\,\left( {{x_{1}}\over {2\,{\sqrt{3}}}} +
     {{a\,\ln  a^2}\over {4\,{\sqrt{3}}}} \right)  \ . \label{diffeq}
\eea

\ni {\bf 3. Symmetry breaking in curved spacetime}.
 We can study now the influence of gravity on the symmetry breaking
pattern of the theory. A simple inspection reveals that a positive
curvature tends to protect the symmetry of the vacuum, and also a
negative one may trigger the breaking of the symmetry.

 In flat spacetime the symmetry is broken whenever either $\lambda_1$
or
$\lambda_2$ are greater than $\frac{8 \pi ^2}{\Lambda ^2}$, that is why
we shall take in the sequel $k_1$ and $k_2$ to be $l_1 - 8 \pi ^2$ and
$l_2 - 8 \pi^2$ respectively. We may illustrate these remarks by
studying the evolution of the
minimum, given by two coordinates (which we choose to be $a$ and $x_2$)
in several circumstances. Before including gravity, it is worth
discussing in more detail the situation in flat space. In fact, several
cases may be analyzed (see \cite{5}). To introduce this study it is
convenient to define
\[ \pm \mu_i ^2 \equiv \Lambda^2 \left( 1- \frac{8
\pi ^2}{\lambda_i ^\pm } \right) \ , \]
where $\lambda_i ^{\pm }$ means $\lambda_i > 8 \pi^2$ (for the $+$ sign)
or $\lambda_i < 8 \pi^2$ (for the $-$ sign). One sticks here to the case
when
$\mu_i \ll \Lambda $. We only repeat the two situations given when
$\mu_2^2 > 3 \mu_1^2$ and ($\lambda_1^+,\,\,\lambda_2^-$), or
$\mu_2^2 < 3 \mu_1^2$ and ($\lambda_1^-,\,\,\lambda_2^+$). In both cases
the results may be summarized by
\bea
\chi_1^2 & = & \frac{\mu_1^2}{\ln \frac{\Lambda^2}{m^2}} \ \left|
1-\frac{3 \mu_1^2}{\mu_2^2} \right|^3 \   \left[ 1 +
{\cal O } \left( \frac{1}{\ln \frac{\Lambda ^2}{m^2}}\right)
\right] \ , \\ \nn
\chi_2 & = & - \sqrt{3} \chi_1 \left( \frac{\mu_1}{\mu_2} \right) ^2
\left[ 1 +
{\cal O } \left( \frac{1}{\ln \frac{\Lambda ^2}{m^2}}\right)
\right] \ , \\ \nn
m^2 & = & \frac{ \mu_1 ^2 \mu_2^2}{\mid \mu_2^2 - 3 \mu_1 ^2 \mid\,\ln
\frac{\Lambda ^2 }{m^2} } \ . \\ \nn
\eea

 As a first example, consider the case in which the coupling constants
are such that the symmetry is not broken in flat space, that is to say,
$\lambda_i < 8 \pi^2$. Let us
see how the situation is modified as we move from negative to positive
curvature. In all the figures, we have represented the values at the
minimum of $a$ and $x_2$, which correspond to the dynamically generated
mass
($m$) and $\chi_2$ transformed to natural units, as given in expression
(\ref{varnat}). We observe that there is a negative value of the
curvature
above which the symmetry is restored (see Figs. 1 and 2).
We observe a continuous phase transition. Some comments about
this fact are done below.

We can see the equivalent plots (Figs. 3 and 4) with
the only difference that one of the couplings is such that the symmetry
is broken at $\R =0$. One can identify a minimum positive value of the
curvature for the symmetry to be restored. The character of
the transition is still continuous.

 This observation may be distressing
if one recalls the results of \cite{8,9}, where the NJL model
was studied under the influence of a gravitational field in different
situations. The conclusion of those works was always that if the
coupling constant
is greater than the critical value in flat space, there is a first
order phase transition at some positive value of the curvature.
In fact there is no actual contradiction. The point is that, till now,
we have
concentrated ourselves in cases where the coupling constants are around
$8 \pi^2$.
We observe that if we explore regions in the space of parameters where
$\lambda_2$ is much smaller, and $\lambda_1$ is greater than $8 \pi^2$
---which would correspond to a limit where our theory approaches the
Gross-Neveu model--- then there is a first order phase transition
for some positive value of the curvature.

Let us now illustrate what happens if we keep the
curvature
constant and vary the values of the couplings, to study how the
domain
which guarantees that the symmetry is not broken is modified by the
presence of curvature. The situation of negative curvature is
exemplified in Figs. 5 and 6. We see that the symmetry is
broken from a negative value of $k_1$, in other words, the region of
parameters of the theory where the symmetric vacuum is stable has
shrinked.

 To finish this discussion we consider the case of positive
curvature. Here one  can see that the aforementioned domain has been
enlarged by the influence of positive curvature (see Figs. 7 and 8).

 In summary, we have calculated the effective potential of composite
scalars in the four-fermion model (\ref{1}) in curved spacetime, and
numerically investigated the phase structure of the theory.

 There is an interesting question about this model ---namely whether it
would be possible for some generalization of it to be represented in
renormalizable form. As we see, if $\lambda_2 = 0$ in (\ref{1}), this is
perfectly feasible as long as one takes into account those terms in the
field $\chi_1$ which transform the theory into a
renormalizable Yukawa-type model which describes near the critical point
the physics of chiral symmetry breaking. As we see, the higher
derivative term in \cite{1} acts against such generalization at
$\lambda_2 \neq 0$.
\bs


\noindent{\large \bf Acknowledgments}

This work has been supported by DGICYT (Spain), project Nos.
PB93-0035
and SAB93-0024, by CIRIT (Generalitat de Catalunya).
S. L. gratefully acknowledges an F.P.I. scholarship from Generalitat de
Catalunya.
\bs

\ni {\bf Appendix: Expansion of the effective potential up to
linear curvature terms}

 In this appendix we give  details of the intermediate steps which are
required in the lengthy calculation of the effective potential in
the linear curvature approximation (for the representation
of propagators in local momentum expansion and weak curvature, see
\cite{10}).
 The first step is the inversion of the operator given by
\bea
\left( i \not\!\nabla -m + 2 \sqrt{3} \chi_2 \frac{\nabar2}{\Lambda ^2}
\right)
\left( i \not\!\nabla +m - 2 \sqrt{3} \chi_2 \frac{\nabar2}{\Lambda ^2}
\right)
=  -\left[ \nabar2 +\left( m-2 \sqrt{3} \chi_2 \nablam
\right)^2 \right] \\ \nn
=-\left[ \left( \Box + \frac{\R}{4} \right) \left( 1-4 \sqrt{3}
\chi_2 \frac{m}{\Lambda ^2} \right) + m ^2+ 12
\frac{{\chi_2}^2}{\Lambda ^4} \left( \Box + \frac{\R}{4} \right) ^2
\right]                      \ ,
 \label{scalope}
\eea
where $\Box$ stands for the spinor d'Alembertian.
Henceforth we shall use in our intermediate calculations, the relation
\bea
{\rm R}_{\alpha \beta \mu \nu} = \frac{{\rm R}}{n \left( n -1 \right) }
\left( g_{\alpha \mu} g_{\beta \nu } - g_{\alpha \nu} g_{\beta \mu }
\right) \ ,
\eea
which is valid in a constant-curvature spacetime. As we are restricting
ourselves to the first nontrivial term in an expansion in the number of
derivatives of the metric tensor, this apparent simplification will not
introduce errors to this order.
 Again, keeping only terms proportional to the curvature, we may write
\beq
\Box = \partial ^2 + \frac{\R }{3 n \left( n-1 \right) }
\left( y^\mu
y^\nu - y^2 \eta^{\mu \nu} \right) \partial_\mu \partial_\nu
 + \frac{\R }{n
\left( n-1 \right) }
{\cal J}^\mu_{\ \lambda} y^\lambda \partial_\mu +
\frac{2}{3} \frac{\R }{n} y^\lambda \partial_\lambda \ , \label{box}
\eeq
being the $y^\mu$  Riemann normal coordinates,
 and
\[
\left( i \not\!\nabla -m + 2 \sqrt{3} \chi_2 \frac{\nabar2}{\Lambda ^2}
\right)
\left( i \not\!\nabla +m - 2 \sqrt{3} \chi_2 \frac{\nabar2}{\Lambda ^2}
\right) = {\cal A}_0 - {\cal A}_2 \ ,
\]
where \[ {\cal A}_0 \equiv - \left[ \partial ^2 \left( 1-4 \sqrt{3}
\frac{
\chi_2 m}{\Lambda ^2} \right) + m^2+ 12 \frac{\chi_2^2}{\Lambda ^4}
\partial ^4 \right] \ ,  \]
 and
\bea
{\cal A}_2 \equiv \R \left\{ \left( 1- 4 \sqrt{3} \frac{\chi_2
m}{\Lambda ^2} \right) \left( \frac{1}{3 n \left( n -1 \right) }
\left( y^\mu
y^\nu - y^2 \eta^{\mu \nu} \right) \partial_\mu \partial_\nu +\frac{
{\cal J}^\mu_{\ \lambda} }{n\left( n -1 \right) } y^\lambda
\partial_\mu+
 \right. \right.\\ \nn
\left. \frac{2}{3 \,n} y^\lambda \partial_\lambda + \frac{1}{4} \right)
+ 12
\frac{\chi_2 ^2}{\Lambda ^4} \left[ \frac{\partial ^2}{2} + \frac{2}{3}
\frac{1}{n \left( n - 1 \right) }
\left( y^\mu
y^\nu - y^2 \eta^{\mu \nu} \right) \partial_\mu \partial_\nu
\partial ^2 + \right. \\ \nn
\left. \left. \frac{2}{n \left( n -1 \right) } {\cal J}^\mu _{\ \nu}
y^\lambda \partial_\mu
\partial ^2 + \frac{2}{3 n} \left( \partial^2 + 2 y ^\lambda \partial
_\lambda \partial^2 \right) \right] \right\} \ . \nn
\eea

 Now we write the equation satisfied by the inverse of the operator
given in (\ref{scalope})
\bea
\left( {\cal A}_0 - {\cal A}_2 + \ldots \right) {\rm G} (y) = \delta (y)
\nn
\eea
 The functional $G$ may be given as an expansion in derivatives of the
metric tensor $ {\rm G} = {\rm G}_0 + {\rm G}_2 + \ldots $, where
\bea
{\cal A}_0 {\rm G}_0 (y) & = & \delta (y), \\ \nn
{\rm G}_2 & = & {\rm G}_0 {\cal A} _2 {\rm G}_0  \ .  \nn
\eea
Let us give some explicit forms of these expressions:
\bea
{\rm G}_0 & = &\int \frac{d^n\,k}{(2 \pi)^n} \frac{
\exp{(-i k y)}}{k^2- \left( m+2 \sqrt{3} \frac{\chi_2}{\Lambda ^2} k ^2
\right)^2}, \\ \nn
{\cal A}_2 {\rm G}_0 & = & \int \frac{d^n\,k}{(2 \pi)^n} \exp{(-i k
y)} \R \left\{ \left( 1- 4 \sqrt{3} \chi_2 \frac{m}{\Lambda ^2} \right)
\left( \frac{1}{3\,n \left( n -1 \right) } \left[ n (n-1) g (k^2) +
 \right. \right. \right.
\\ \nn
 & & \left. 2 \left( 2n +3 \right) k^2 g' (k^2) +
4 \left( k^2 \right) ^2 g''(k^2) - \left( 4 k^2 f''(k^2)
+ 2 n f'(k^2) \right) \right]  \\ \nn
& & \left. - \frac{2}{3 n} \left( n\,g(k^2) +
   2 k ^2 g(k^2) \right) + \frac{1}{4} \right) +
12 \frac{\chi_2
^2}{\Lambda ^4} \left[ \frac{1}{2} \bar{g} (k^2) +
\right.   \\ \nn
& &  \frac{2}{3 n (n-1)}
\left[ n(n+1) \bar{g} (k^2)
  + 2 k^2 (2 n +3)\bar{g'} (k^2) + 4 k^2 \bar{g''} -
\left(
4 k^2 \bar{f''}(k^2) +  \right. \right. \\ \nn
& & \left. \left. \left. \left. 2 n \bar{f'} (k^2) \right) \right]
 + \frac{2}{3 n} \left( \bar{g} (k^2) - 2 n \left( \bar{g}
 (k^2)+2 k^2 \bar{g}(k^2) \right) \right) \right] \right\} \ .
\label{verylong}
\eea
Some definitions are in order:
\bea
g(x) & \equiv & \frac{1}{x-\left( a+ 2 \sqrt{3} \chi_2
\frac{x}{\Lambda^2} \right) ^2} \ , \\ \nn
f(x) & \equiv & \frac{x}{x-\left( a+ 2 \sqrt{3} \chi_2
\frac{x}{\Lambda^2} \right) ^2} \ , \\ \nn
\bar{f}(x) & \equiv & \frac{-x^2}{x-\left( a+ 2 \sqrt{3} \chi_2
\frac{x}{\Lambda^2} \right) ^2}  \ , \\ \nn
\bar{g}(x) & \equiv & \frac{-x}{x-\left( a+ 2 \sqrt{3} \chi_2
\frac{x}{\Lambda^2} \right) ^2} \ . \nn
\eea

 Now ${\cal G} = {\cal B} {\rm G}$, where ${\cal B} \equiv
i \not\!\nabla +a - 2 \sqrt{3} \chi_2 \nablap $,
and we may expand ${\cal B} = {\cal B}_0 + {\cal B}_2 + \ldots $ (again
according
to the number of derivatives of the metric tensor). The
explicit expression of the operators ${\cal B}_0$ and ${\cal B}_2 $ are:
\bea
{\cal B}_0 & = & i \not\!\hat{\partial} +a - 2 \sqrt{3} \chi_2
\frac{\partial ^2}{\Lambda ^2} \ , \\ \nn
{\cal B}_2 & = & \frac{i \R}{6 n (n-1) } \left( y ^2
\not\!\hat{\partial}
-\not\!\hat{y} y^\lambda \partial_\lambda \right) + \frac{i \R}{2 n
(n-1)} y_a \hat{\gamma_b} {\cal J}^{a b} \\ \nn
& & - 2 \sqrt{3} \chi_2 \frac{\R}{\Lambda ^2} \left( \frac{1}{3 n (n-1)}
\left(
y^\mu y^\nu - y ^2 \eta^{\mu \nu} \right) \partial_\mu \partial_\nu +
\frac{{\cal J}^\mu_\nu}{n (n-1)} y^\lambda \partial_\mu \right. \\ \nn
 & & \left. +\frac{2}{3 n } y^\lambda \partial_\lambda + \frac{1}{4}
\right) \ ,   \label{bescal}
\eea
where the symbol $\hat{~}$ means that the Dirac matrices involved
satisfy the anticommutation relations $\{ \hat{\gamma}^a  , \
\hat{\gamma}^b \} = \eta^{a b} \ .$

 We need only to expand the expressions shown below
\bea
{\cal G}_0 & = & {\cal B}_0 {\rm G}_0 \ ,  \label{g0cal} \\
{\cal G}_2 & = & {\cal B}_2 {\rm G}_0 + {\cal B}_0 {\rm G}_0 {\cal A}_2
{\rm G}_0 \ .                                \label{g2cal}
\eea
In particular,  the explicit form of ${\cal G}_0$ and ${\cal
B}_2 {\rm G}_0$ are
\bea
{\cal G}_0 & = & \int \frac{d ^n k}{(2 \pi )^n} \frac{
\exp{( -i k y) }}{\not\!\hat{k} - \left( a+ 2 \sqrt{3}
\frac{\chi_2}{\Lambda^2} k^2 \right) ^2} \ ,  \\
{\cal B}_2 {\rm G}_0 & = & - \int \frac{d ^n k}{( 2 \pi ) ^n}
\exp{ (-i k y) } 2 \sqrt{3} \chi _2 \frac{\R}{\Lambda^2} \left(
\frac{1}{3 n (n-1)} {\rm h} 1 (k^2)- \frac{2}{3 n} {\rm h} 2 (k^2) +
\frac{1}{4} \right) \ . \label{falsa}
\eea
Equation (\ref{falsa}) is true,  apart from terms which have an odd
number
of $\hat\gamma$ matrices ---and which will not contribute to the
effective
potential. Finally, ${\rm h}1$ and ${\rm h}2$ are defined as:
\bea
{\rm h}1(k^2) & = & n(n+1) g(k^2) + 2 k^2 (2n +3) g'(k^2)+ \\ \nn
& & + 4 \left( k^2 \right) ^2 g''(k ^2) - \left( 4 k ^2 f''(k^2) + 2 n
f'(k^2) \right) \ , \\ \nn
{\rm h}2(k^2) & = & n g(k^2) + 2 k^2 g'(k^2) \ .
\eea

 Putting all these pieces together we may construct the propagator
${\cal G}$ (apart from terms with an odd number of $\hat{\gamma}$
matrices) and perform the momentum integrations ensuing from the Tr
operations in (\ref{deriv1}) and (\ref{deriv2}). As for ({\ref{deriv2}),
remember that $\not\!\nabla ^2 = \Box + \frac{\R}{4}$ and $\Box $ is
given in (\ref{box}).

\newpage

\begin{center}
{\Large \bf Figure captions}
\end{center}

\vskip1cm
\ni{\bf Fig. 1}
The coupling constants are kept constant at the values
given by $k_1=-2=k_2$. One sees a continuous phase transition.

\vskip0.5cm

\ni{\bf Fig. 2}
The coupling constants are kept constant at the values
given by $k_1=-2=k_2$. There is a continuous phase transition at
some negative value of the curvature.

\vskip0.5cm

\ni{\bf Fig. 3}
The coupling constants are kept constant at the values
given by $k_1=-1, \ k_2=0.2$. Now we see a continuous phase
transition.

\vskip0.5cm

\ni{\bf Fig. 4}
The coupling constants are kept constant at the values
given by $k_1=-1 \ k_2=0.2$.

\vskip0.5cm

\ni{\bf Fig. 5}
The curvature is such that $r=-0.0008$ and $k_2$ is kept constant
at $k_2=-0.8$.

\vskip0.5cm
\ni{\bf Fig. 6}
The curvature is such that $r=-0.0008$ and  $k_2$ is kept constant
at $k_2=-0.8$.

\vskip0.5cm

\ni{\bf Fig. 7}
The curvature is such that $r=0.0008$ and $k_2$ is kept
constant
at $k_2=-0.8$. Now the range of values of the coupling constants which
maintain the symmetry of the vacuum is enlarged as the curvature gets
positive.

\vskip0.5cm

\ni{\bf Fig. 8}
The curvature is such that $r=0.0008$ and $k_2$ is kept constant
at $k_2=-0.8$.

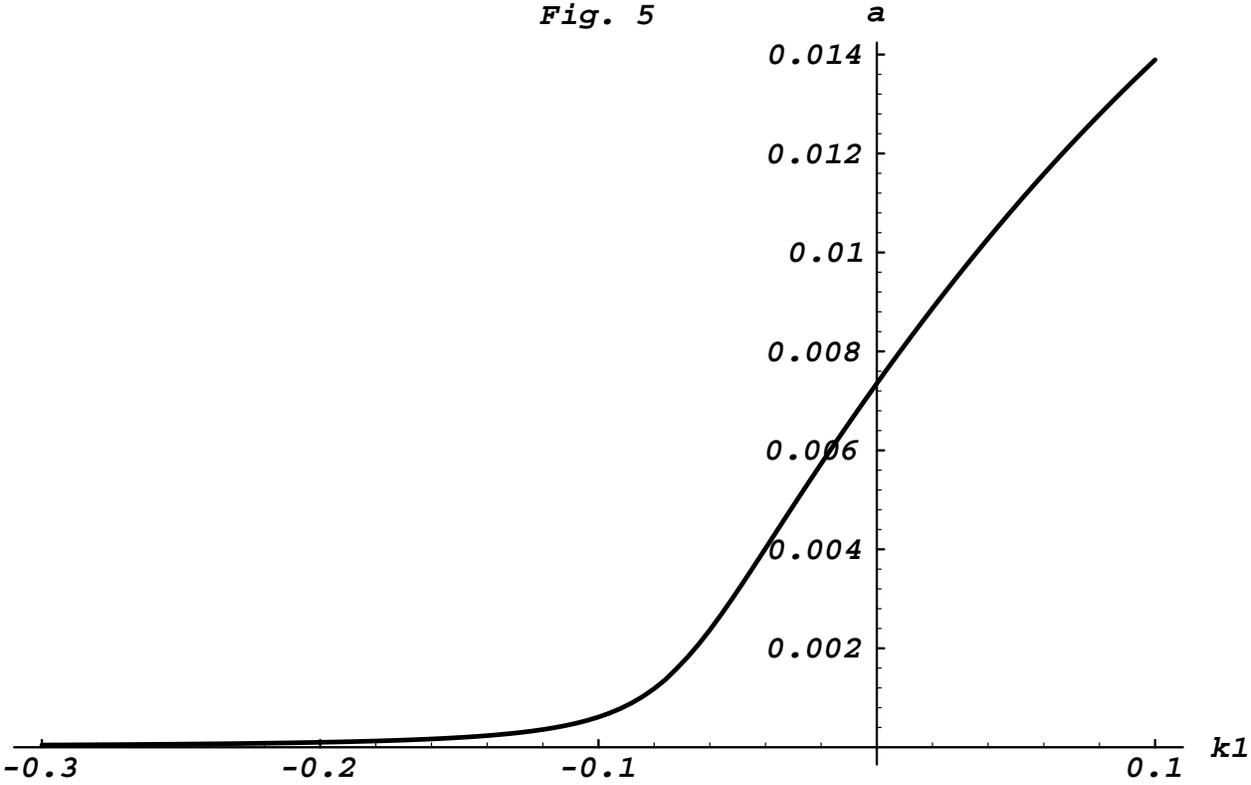
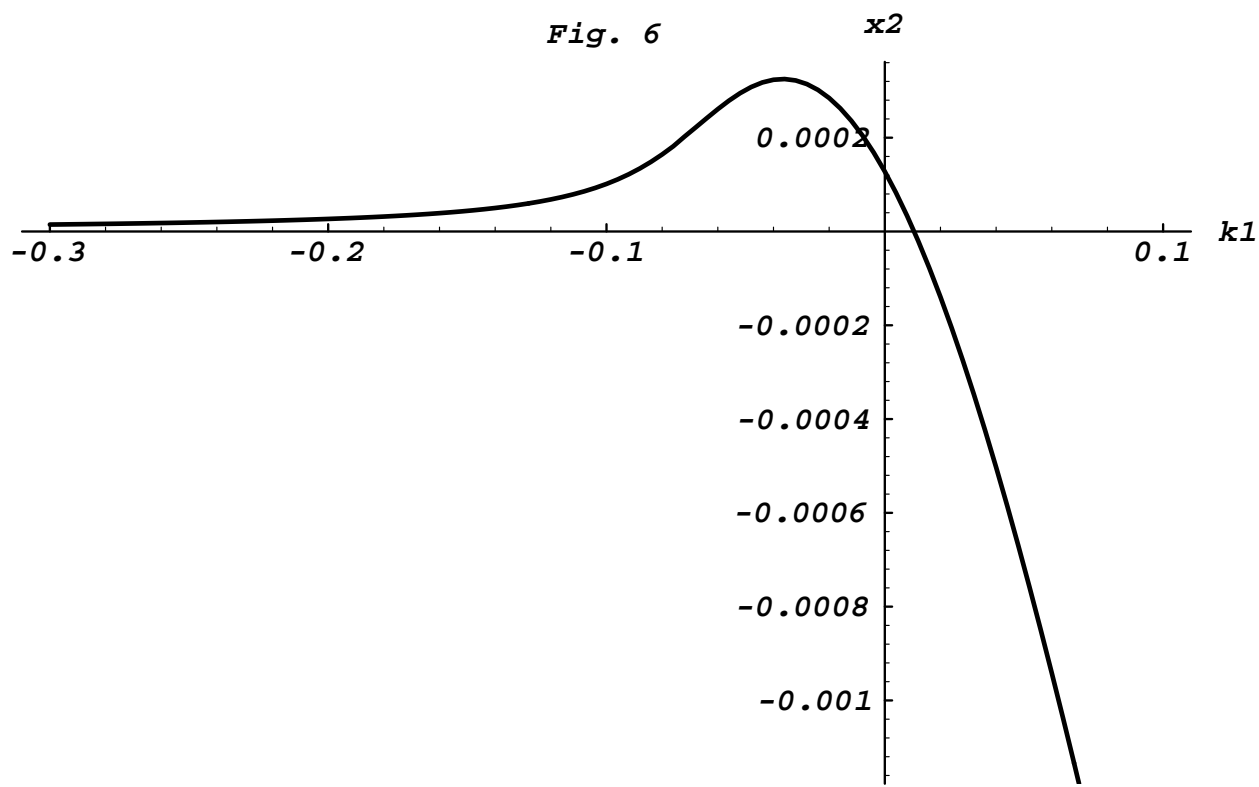
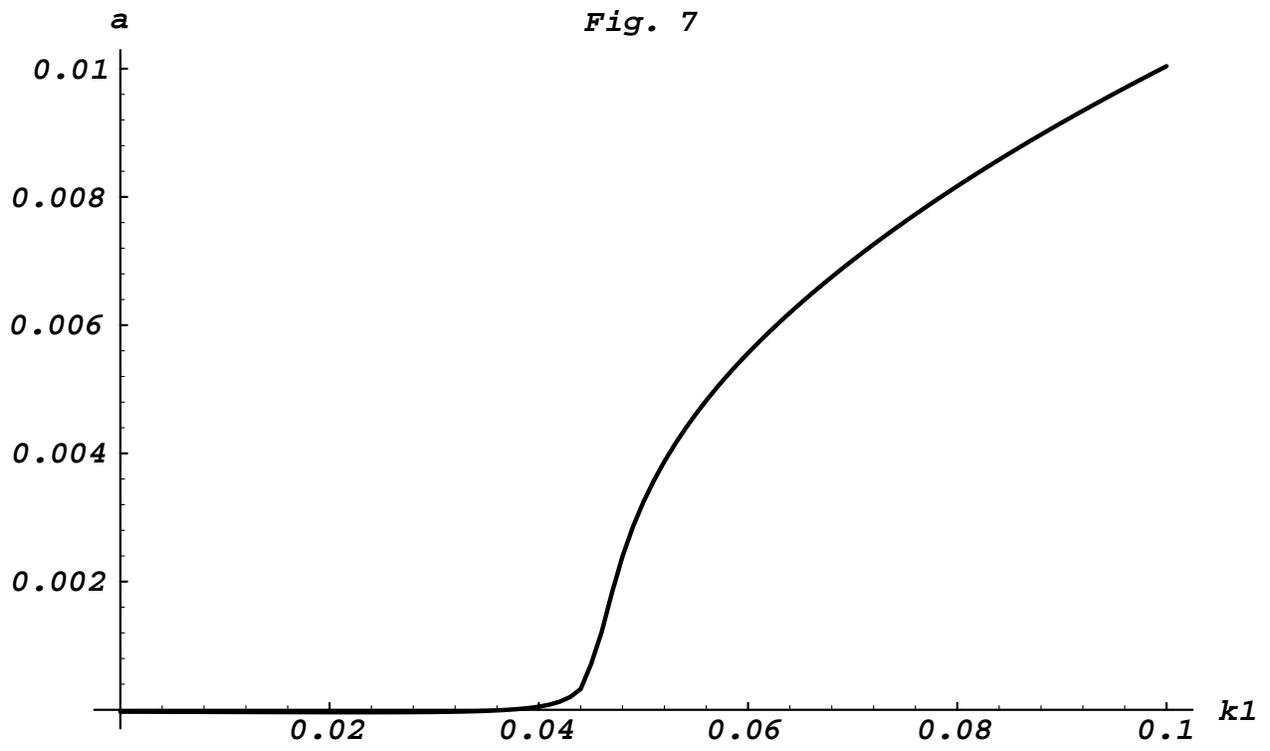
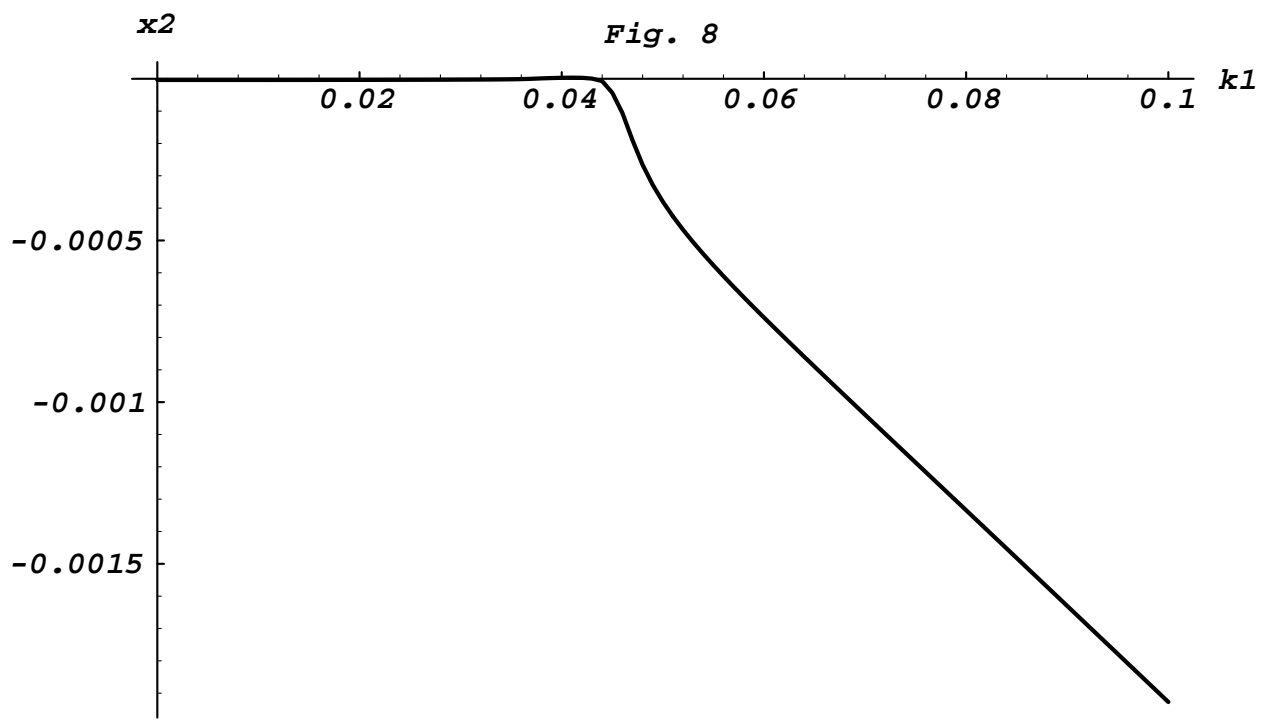
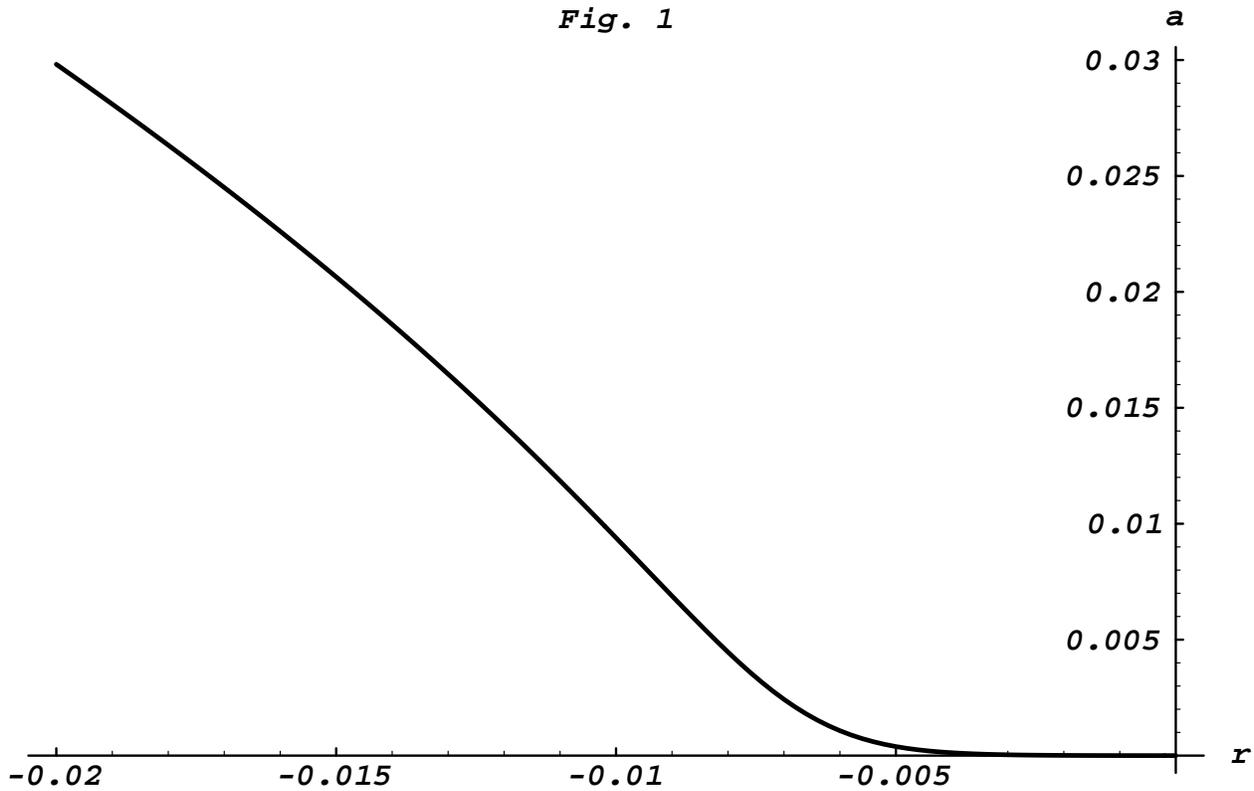
\end{document}